\documentclass[showkeys,showpacs,superscriptaddress]{revtex4}
\usepackage{amsmath}
\usepackage{amssymb}
\DeclareMathOperator\arctanh{arctanh}

\DeclareMathOperator{\sech}{sech}

\begin{document}

\title{
Spinor field solutions in $F\left(B^2\right)$ modified Weyl gravity
}

\author{
Vladimir Dzhunushaliev
}
\email{v.dzhunushaliev@gmail.com}
\affiliation{
Department of Theoretical and Nuclear Physics,  Al-Farabi Kazakh National University, Almaty 050040, Kazakhstan
}
\affiliation{
Institute of Experimental and Theoretical Physics,  Al-Farabi Kazakh National University, Almaty 050040, Kazakhstan
}
\affiliation{
National Nanotechnology Laboratory of Open Type,  Al-Farabi Kazakh National University, Almaty 050040, Kazakhstan
}
\affiliation{
Academician J.~Jeenbaev Institute of Physics of the NAS of the Kyrgyz Republic, 265 a, Chui Street, Bishkek 720071, Kyrgyzstan
}

\author{Vladimir Folomeev}
\email{vfolomeev@mail.ru}
\affiliation{
Institute of Experimental and Theoretical Physics,  Al-Farabi Kazakh National University, Almaty 050040, Kazakhstan
}
\affiliation{
National Nanotechnology Laboratory of Open Type,  Al-Farabi Kazakh National University, Almaty 050040, Kazakhstan
}
\affiliation{
Academician J.~Jeenbaev Institute of Physics of the NAS of the Kyrgyz Republic, 265 a, Chui Street, Bishkek 720071, Kyrgyzstan
}

\affiliation{
International Laboratory for Theoretical Cosmology, Tomsk State University of Control Systems and Radioelectronics (TUSUR),
Tomsk 634050, Russia
}

\begin{abstract}
We consider modified Weyl gravity where a Dirac spinor field is nonminimally coupled to gravity.
It is assumed that such modified gravity is some approximation for the description of quantum gravitational effects
related to the gravitating spinor field.
It is shown that such a theory contains solutions
for a class of metrics which are conformally equivalent to the Hopf metric on the Hopf fibration. For this case, we obtain
a full discrete spectrum of the solutions and show that they can be related to the Hopf invariant on the Hopf fibration.
The expression for the spin operator in the Hopf coordinates is obtained. It is demonstrated that this class of conformally
equivalent metrics contains: (a) a metric describing a toroidal wormhole without exotic matter;
(b) a cosmological solution with a bounce and inflation;
and (c) a transition with a change in metric signature.
A physical  discussion of the results is given.
\end{abstract}



\date{\today}

\maketitle

\section{Introduction}

The solving of general relativistic equations sourced by spinor fields offers great difficulties.
Apparently, this is caused by the fact that general relativity  (GR) is a purely classical theory, while spinor fields are described by the Dirac equation
that is used for describing  purely quantum particles. This problem appears to be indirectly related to the problems arising while quantising GR.
Consistent with all of this, one may conclude that deriving solutions in gravity supported by spinor fields can help one to understand what is the quantization of gravity.

At the present time the literature in the field offers spherically symmetric asymptotically flat spinor field solutions within GR
\cite{Finster:1998ws,Finster:1998ux,Herdeiro:2017fhv,Dzhunushaliev:2018jhj,Dzhunushaliev:2019kiy},
but such solutions always involve {\it two} spinor fields having opposite spins, so that their total spin is zero. Therefore the main difficulty with deriving solutions supported
by spinor fields is to find solutions with {\it one} spinor field whose spin is already nonzero.
In the present paper we seek possible ways to modify GR to get solutions with one spinor field.  We will show that this can be done within modified Weyl gravity
where a nonminimal interaction between matter and gravity is introduced. This type of interaction
has been repeatedly investigated in the literature in various aspects
(see, e.g., Refs.~\cite{Sushkov:2009hk,Tamanini:2013aca,Dzhunushaliev:2013nea,Harko:2014gwa,Dzhunushaliev:2015mva} and references inside).

Conformal Weyl gravity is one of numerous ways to generalize GR. Since GR is a purely classical theory, it
does not take quantum effects into account.
To involve them into consideration, it is necessary to go beyond the Einstein theory.
In the simplest case, the modification of GR reduces to the replacement of the Einstein gravitational Lagrangian $\sim R$ by the
modified Lagrangian $\sim f(R)$, where $f(R)$ is some function of the scalar curvature $R$.
Such  modified gravities had initially been applied for the description of
the evolution of the very early Universe, but it was shown in recent years that they can also be successfully applied to model  various
cosmological aspects of the present Universe (for a general review on the subject, see, e.g., Refs.~\cite{Nojiri:2010wj,Nojiri:2017ncd}).

For its part, in conformal Weyl gravity, the scalar curvature $R$ in the gravitational action is replaced by a scalar invariant
containing the Weyl tensor.
From the point of view of quantum theory, an important advantage of Weyl gravity is that it is conformally invariant and
renormalizable~\cite{Mannheim:2009qi}. The main weakness of this theory is that its viability is questionable, since
it does not satisfy some tests in the Solar System~\cite{Flanagan:2006ra}. Nevertheless, this theory is quite widely used in the literature both in modelling
the observed galactic rotation curves without involving dark matter~\cite{Mannheim:1988dj,Mannheim:2010xw} and in solving cosmological puzzles related
to the problem of the explanation of the accelerated expansion of the present Universe~\cite{Mannheim:2005bfa,Diaferio:2011kc}.

Consistent with this, we will work here within modified Weyl gravity,
which enables us to get self-consistent solutions supported by a gravitating spinor field.
As we will see below, this can be achieved by introducing a nonminimal interaction between matter and gravity. The main idea is to find such a form of the interaction that, in some cases, would lead
to the automatic fulfilment of gravitational equations. It turns out that it is possible if one considers a spacetime metric whose spatial part is conformally equivalent to the Hopf fibration
on which one can find nontrivial solutions to the Dirac equation.
In this case, the left-hand side of the Bach equation (i.e., the Bach tensor) vanishes identically.  Then one can describe the modified interaction between matter and gravity
by introducing some coefficient depending on the Bach scalar curvature. If this coefficient is chosen so that it is equal to zero for vanishing Bach tensor,
then it is evident that the gravitational Bach equation will identically satisfied.
Thus, the purpose of the current study is to modify Weyl gravity in such a manner as to obtain a self-consistent system describing a nonminimal interaction between a gravitational field and
{\it one} spinor field.

It should be emphasized that modified Weyl gravity under investigation, which involves the nonminimal coupling to matter,  is regarded as
\textit{an approximate way of describing quantum gravitational effects emerging when gravity is coupled to a spinor field}.
This means that in strong gravitational fields created by the spinor field it is necessary to take into account quantum gravitational effects
which are assumed to be approximately described by using modified Weyl gravity. In doing so, in the context under consideration,
we suppose that modified Weyl gravity is not a fundamental theory, but it is used only for an approximate description of the aforementioned quantum effects.

The paper is organized as follows. In Sec.~\ref{gen_eqs}, we present general equations for the system under consideration. In Sec.~\ref{solutions},
we choose a special form of the metric that enables us (i) to make the Bach tensor equal to zero; (ii) to ensure that the gravitational equations are automatically satisfied;
and (iii) to solve the Dirac equation analytically. In Sec.~\ref{worm_sol},
we demonstrate the possibility of obtaining various solutions within Weyl gravity under consideration,
and in Sec.~\ref{spin_oper} we obtain the expression for the spin operator in the Hopf coordinates. Finally, in Sec.~\ref{concl}, we summarize and discuss the results obtained.

\section{General equations of $F(B^2)$ modified Weyl gravity}
\label{gen_eqs}

The action for conformal Weyl gravity minimally coupled to a spinor field can be written in the form
(hereafter we work in natural units $\hbar = c = 1$)
\begin{equation}
	\mathcal S = \int d^4 x \sqrt{-g} \left(
		- \alpha_g C_{\alpha \beta \gamma \delta}
		C^{\alpha \beta \gamma \delta} + \mathcal L_\psi
		\right) ,
\label{Weyl_10}
\end{equation}
where $\alpha_g$ is a dimensionless constant,
$
	C_{\alpha \beta \gamma \delta} = R_{\alpha \beta \gamma \delta} +
	\frac{1}{2} \left(
		R_{\alpha \delta} g_{\beta \gamma} -
		R_{\alpha \gamma} g_{\beta \delta} +
		R_{\beta \gamma} g_{\alpha \delta} -
		R_{\gamma \delta} g_{\alpha \beta}
	\right)
$ is the Weyl tensor, and  $L_\psi $ is the Lagrangian density of the massless spinor field,
$$
	\mathcal L_\psi = \frac{i }{2} \left(
			\bar \psi \gamma^\mu \psi_{; \mu} -
			\bar \psi_{; \mu} \gamma^\mu \psi
		\right),
$$
which contains the covariant derivative
$
\psi_{; \mu} \equiv \left[\partial_{ \mu} +1/8\, \omega_{a b \mu}\left(
\gamma^a  \gamma^b- \gamma^b  \gamma^a\right)\right]\psi
$,
where $\gamma^a$ are the Dirac matrices in
flat space (below we use the spinor representation of the matrices). In turn, the Dirac matrices in curved space,
$\gamma^\mu = e_a^{\phantom{a} \mu} \gamma^a$, are obtained using
the tetrad
$ e_a^{\phantom{a} \mu}$, and $\omega_{a b \mu}$ is the spin connection
[for its definition, see Ref.~\cite{Lawrie2002}, formula (7.135)].

The action \eqref{Weyl_10} and hence the corresponding theory are invariant under the conformal transformations
$$
	g_{\mu \nu} \rightarrow f(x^\alpha) g_{\mu \nu}, \quad
	\psi \rightarrow f^{-3/2}(x^\alpha) \psi ,
$$
where the function $f(x^\alpha)$ is arbitrary.

To the best of  our knowledge, efforts to find spherically and axially symmetric solutions supported by {\it one} spinor field were unsuccessful
in Weyl gravity, although such solutions do exist in GR (see, e.g., Ref.~\cite{Bronnikov:2019nqa}).
It seems to us that this is a consequence of the fact that, from the physical point of view, both the GR Lagrangian and the Lagrangian \eqref{Weyl_10} describe the interplay of a classical theory (either Einstein or Weyl gravities) and a purely quantum theory (the Dirac theory describing an electron with the spin). In some sense, one can say that this problem may be indirectly related to the problem of quantization of gravity. In this connection, the question arises: is it somehow possible to modify gravity in such a manner as to obtain solutions supported by a spinor field within such theories of gravity?

In order to obtain solutions with one spinor field, we consider the following modified Weyl gravity:
\begin{equation}
	\mathcal L_{mW} = - \alpha_g C_{\alpha \beta \gamma \delta}
	C^{\alpha \beta \gamma \delta} +
	F \left( B^2 \right) \mathcal L_\psi ,
\label{Weyl_50}
\end{equation}
where $B^2 = B_{\mu \nu} B^{\mu \nu}$ is the Bach scalar curvature invariant,
$
	B_{\mu \nu} =
	C^{\phantom{\mu}\alpha \phantom{\nu} \beta}_
	{\mu \phantom{\alpha}\nu \phantom{\beta} ; \alpha \beta} -
	\frac{1}{2}
	C^{\phantom{\mu}\alpha \phantom{\nu} \beta}_
	{\mu \phantom{\alpha}\nu \phantom{\beta} } R_{\alpha \beta}
$ is the Bach tensor,
  and
 $F \left( B^2 \right)$ is an arbitrary function.

Within such a theory, after the variation of the corresponding action with respect to the metric and spinor field, the field equations take the form
\begin{eqnarray}
	- 2 \alpha_g B_{\mu \nu} +
	\frac{1}{\sqrt{-g}}
	\frac{\delta}{\delta g^{\mu \nu}} \left[
		\sqrt{-g} F \left( B^2 \right)
	\right] \mathcal L_\psi  &=&
	- \frac{1}{2} F \left( B^2 \right) T_{\mu \nu} ,
\label{Weyl_60}\\
	i \gamma^\mu \psi_{;\mu} &=& 0 ,
\label{Weyl_70}
\end{eqnarray}
where the spinor field energy-momentum tensor  $T_{\mu \nu}$ is defined as
\begin{equation}
	T_{\mu \nu} = \frac{i}{4} \left[
		\bar\psi \gamma_{\mu} \psi_{; \nu} +
		\bar\psi \gamma_\nu \psi_{;\mu} 	-
		\bar\psi_{; \mu}\gamma_{\nu} \psi -
		\bar\psi_{; \nu} \gamma_\mu \psi
	\right] - g_{\mu \nu} \mathcal L_{\psi} .
\label{Weyl_80}
\end{equation}

As usual, by the Dirac equation \eqref{Weyl_70},
the spinor field Lagrangian $\mathcal L_\psi=0$
and hence the second term in the left-hand side of the modified Bach equation~\eqref{Weyl_60}
and the last term in Eq.~\eqref{Weyl_80} vanish. As a result,
we have the following set of equations in $F \left( B^2 \right)$ modified Weyl gravity:
\begin{eqnarray}
	\alpha_g B_{\mu \nu} &=&
	\frac{1}{4 \alpha_g} F \left( B^2 \right) T_{\mu \nu} ,
\label{Weyl_90}\\
	i \gamma^\mu \psi_{;\mu} &=& 0 ,
\label{Weyl_100}\\
	T_{\mu \nu} &=& \frac{i}{4} \left[
		\bar\psi \gamma_{\mu} \psi_{; \nu} +
		\bar\psi \gamma_\nu \psi_{;\mu} 	-
		\bar\psi_{; \mu}\gamma_{\nu} \psi -
		\bar\psi_{; \nu} \gamma_\mu \psi
	\right] .
\label{Weyl_110}
\end{eqnarray}


\section{The results of calculations}
\label{solutions}

To find a solution of Eqs.~\eqref{Weyl_90} and \eqref{Weyl_100}, we choose the metric in the form
\begin{equation}
	ds^2 = f^2 \left(
		t, \chi, \theta, \varphi
	\right) \left\{
		dt^2 - \frac{r^2}{4}
		\left[
			\left( d \chi - \cos\theta d \varphi\right)^2
			+ d \theta^2 + \sin^2 \theta d \varphi^2
		\right]
	\right\} = f^2 \left(
	t, \chi, \theta, \varphi
	\right) \left( dt^2 - r^2 dS^2_3 \right).
\label{solution_10}
\end{equation}
Here $f \left( t, \chi, \theta, \varphi \right)$ is an arbitrary function,
$dS^2_3$ is the Hopf metric \eqref{A_130} on the unit sphere
(for a detailed description of the Hopf fibration, see Appendix~\ref{Hopf_bundle}), and $r=\text{const}$. For the metric~\eqref{solution_10}, the Bach tensor
$$
	B_{\mu \nu} = 0 .
$$
This enables us to simplify considerably the equations of $F \left( B^2 \right)$ modified Weyl gravity. Namely, by choosing
$$
	F \left( B^2 \right) = 0 ,
$$
the modified Bach equation~\eqref{Weyl_90} is identically satisfied and the only remaining equation is the Dirac equation~\eqref{Weyl_100}.

In fact, this is some trick of solving the Bach equation \eqref{Weyl_90}: we choose the nonminimal coupling $F \left( B^2 \right)$
in such a manner as to provide the automatic fulfilment of this equation. But it should be mentioned that this trick is very nontrivial;
for instance, it is apparently impossible to get similar solution in GR. The reason is that in GR such a nonminimal coupling would look like
$F\left(R_{\mu \nu} R^{\mu \nu} \right) $, and to obtain the required solution one has to have a metric which gives
$R_{\mu \nu} = 0$. If one seeks a solution in the class of asymptotically flat metrics, such solutions are Schwarzschild and Kerr black holes.
But for these cases the condition $R_{\mu \nu} = 0$ is satisfied everywhere  except the singularity where the Ricci invariant
$R_{\mu \nu} R^{\mu \nu}$ diverges.

\subsection{Dirac equation}

To solve Eq.~\eqref{Weyl_100}, we use the following
$\mathfrak{Ansatz}$ for the spinor field:
\begin{equation}
	\psi_{m,n} =f^{-3/2} e^{-i \tilde \Omega t} e^{i n \chi} e^{i m \varphi}
	\begin{pmatrix}
		\Theta_1(\theta)  \\
		\Theta_2(\theta)  \\
		\Theta_3(\theta)  \\
		\Theta_4(\theta)
	\end{pmatrix} .
\label{3_50}
\end{equation}
This spinor may transform under a rotation through an angle
$2 \pi$  as follows:
\begin{equation}
	\psi_{m,n}(\chi + 2 \pi, \theta, \varphi + 2 \pi) = \pm
	\psi_{m,n}(\chi , \theta, \varphi).
\label{3_60}
\end{equation}
Taking into account the factors $e^{i n \chi}$ and $e^{i m \varphi}$ from
Eq.~\eqref{3_50}, one can see that the numbers $m$ and $n$ should satisfy the condition
$$
	m + n = \frac{N}{2},
$$
where $N = 0, \pm 1, \pm 2 , \dots$ .

To solve the Dirac equation, we employ the tetrad
$$
  e^a_{\phantom{a} \mu} =
  f \left( t, \chi, \theta, \varphi \right)
  \begin{pmatrix}
    1 & 		0 		& 0 			& 0 						\\
    0 & 	\frac{r}{2}	& 0 			& - \frac{r}{2} \cos \theta 	\\
    0 & 0				& \frac{r}{2} 	& 0 						\\
    0 & 0				& 0 			& \frac{r}{2} \sin \theta	
  \end{pmatrix}
$$
coming from the metric \eqref{solution_10}.

Upon substituting the $\mathfrak{Ansatz}$~\eqref{3_50} into
Eq.~\eqref{Weyl_100}, we have four equations
\begin{eqnarray}
	\Theta_{1,3}^\prime + \Theta_{1,3} \left(
		\frac{\cot \theta}{2} + n
	\right) +
	\Theta_{2,4} \left(
		\frac{1}{4} \mp \frac{\Omega}{2} - n \cot \theta -
		\frac{m}{\sin \theta}
	\right) &=& 0 ,
\label{3_80}\\
	\Theta_{2,4}^\prime + \Theta_{2,4} \left(
		\frac{\cot \theta}{2} - n
	\right) + \Theta_{1,3}  \left(
		- \frac{1}{4} \pm \frac{\Omega}{2} - n \cot \theta - \frac{m}{\sin
	\theta}
	\right) &=& 0 ,
\label{3_90}
\end{eqnarray}
where the dimensionless $\Omega = r \tilde \Omega$. In these equations, the upper sign is used for the functions
$\Theta_{1}$ and $\Theta_{2}$, and the lower~-- for $\Theta_{3}$ and $\Theta_{4}$; the prime denotes differentiation with respect to $\theta$.

\subsection{The solution of the Dirac equation}

It is evident that the set of equations \eqref{3_80} and \eqref{3_90} is split into two set of equations: one for the unknown functions
$\Theta_{1,2}$ and the other one -- for the functions $\Theta_{3,4}$. The corresponding particular solutions can be represented in the form
\begin{equation}
	\left( \Theta_{1,3} \right)_{m,n} =
	\pm \left( \Theta_{2,4} \right)_{m,n} =
	C \sin^\alpha \left( \frac{\theta}{2} \right)
	\cos^\beta \left( \frac{\theta}{2} \right) ,
\label{3_a_10}
\end{equation}
where
$$	\alpha = \pm\left(n+m\right) - \frac{1}{2}, \quad
	\beta = \pm \left(n-m\right) - \frac{1}{2} ,
\quad
	\Omega_n = \frac{\epsilon}{2}\left(1\pm 4 n\right)
$$
and $C$ is a complex integration constant; the parameter $\epsilon=1$ and $-1$ for the pairs $\Theta_{1,2}$ and $\Theta_{3,4}$, respectively.
These solutions form a discrete spectrum depending on two quantum numbers $m$ and $n$.

Let us now represent the solution \eqref{3_a_10} for different possible relations between the numbers $\alpha$ and $\beta$:
\begin{itemize}
	\item In the case of $\alpha > \beta$, we have
	$$
		\left( \Theta_{1,3} \right)_{m,n} =
		\pm \left( \Theta_{2,4} \right)_{m,n} =
		\tilde C \sin^{\alpha - \beta} \left( \frac{\theta}{2} \right)
		\sin^\beta \theta .
	$$
	\item In the case of $\alpha = \beta$,
	$$
	\left( \Theta_{1,3} \right)_{m,n} =
	\pm \left( \Theta_{2,4} \right)_{m,n} =
	\tilde C \sin^\beta \theta .
	$$
	\item In the case of $\alpha < \beta$,
	$$
		\left( \Theta_{1,3} \right)_{m,n} =
		\pm \left( \Theta_{2,4} \right)_{m,n} =
		\tilde C \cos^{\beta - \alpha} \left( \frac{\theta}{2} \right)
		\sin^\alpha \theta .
	$$
\end{itemize}
In the above expressions,  $\tilde C$ is a rescaled integration constant.


\subsection{
A check of the orthogonality condition for the discrete spectrum of the solutions $\psi_{m,n}$
}

The scalar product of the eigenfunctions $\psi_{m, n}$ of the discrete spectrum of the solutions to the Dirac equations is
$$
	\left\langle \psi_{m, n} | \psi_{p, q} \right\rangle =
	\int \psi^{\dag}_{m, n} \psi_{p, q} d V =
	\frac{\pi^2 r^3}{2} \int \limits_0^\pi
	\sin \theta \left(
		\Theta_1^2 + \Theta_2^2 + \Theta_3^2 + \Theta_4^2
	\right) d \theta ,
$$
 where $d V = \sqrt{\gamma} d \chi d \theta d \varphi$ and $\gamma$ is the determinant of the three-dimensional part of the metric~\eqref{solution_10}.
Using the solutions \eqref{3_a_10}, this formula yields for both pairs of the functions
$$
	\left\langle \psi_{m, n} | \psi_{p, q} \right\rangle =
	2 \left| C \right|^2 \pi^2 r^3
	\int \limits_0^\pi \sin \theta \sin^{2 \alpha} \left( \frac{\theta}{2} \right)
	\cos^{2 \beta} \left( \frac{\theta}{2} \right) d \theta =
	4 \left| C \right|^2\pi^2 r^3
	\frac{\Gamma(1 + \alpha) \Gamma(1 + \beta)}
	{\Gamma(2 + \alpha + \beta)} , \quad
	\alpha, \beta > - 1 .
$$
It is seen from this expression that the functions $\psi_{m, n}$ and $\psi_{p, q}$ are not orthogonal to each other.

To construct a set of orthogonal functions, one can use the Gram-Schmidt process that is a method for orthonormalizing a set of vectors in an inner product space.
This can be done in two stages:
\begin{enumerate}
	\item Since we have an infinitely dimensional space of the wave functions, the first step is to renumber all the functions $\psi_{m, n}$.
This can be done, since the set of indices $m$ and $n$ is a denumerable one.
	\item The second step is to employ the well-known Gram-Schmidt process to obtain an orthonormal set of the functions $\psi_{m, n}$.
\end{enumerate}

\subsection{Current density}

The current density is defined by the expression
$$
	j_\mu = \bar \psi \gamma_\mu \psi =
	\frac{r}{f(t, \chi ,\theta ,\varphi )^2}
	\begin{Bmatrix}
		\frac{\Sigma_1}{r} , &	
		- \Sigma_3 ,	&
		0	, &	
		- \frac{1}{2} \left[
			\sin \theta \Sigma_2 - 2 \cos \theta \Sigma_3
		\right]
	\end{Bmatrix} ,
$$
where $
	\Sigma_1 = \Theta_1^2 + \Theta_2^2 + \Theta_3^2 + \Theta_4^2 ,\,
	\Sigma_2 = \Theta_1^2 - \Theta_2^2 - \Theta_3^2 + \Theta_4^2 $, and
$	
\Sigma_3 = \Theta_1 \Theta_2 -  \Theta_3 \Theta_4
$.

Consider the case of $\Theta_1 = \Theta_2 = \Theta$ and $\Theta_3 = \Theta_4 = 0$. Then
$$
	j_\mu = \frac{2  \Theta^2(\theta)}{f(t, \chi ,\theta ,\varphi )^2}
	\begin{Bmatrix}
		1 , &	- \frac{r}{2} ,	& 0	, &	\frac{r}{2} \cos \theta
	\end{Bmatrix}.
$$
Using the 1-form $\Upsilon$ defined in Eq.~\eqref{A_160}, the 1-form $j_\mu d x^\mu$ can be written as
$$
  j_\mu d x^\mu = \frac{2  \Theta^2(\theta)}{f(t, \chi ,\theta ,\varphi )^2}
  \left( dt - \frac{r}{2} \Upsilon \right) .
$$
This means that it is possible to relate the conformally invariant metric~\eqref{solution_10} with the Hopf invariant~\eqref{A_150}, which is $H = 1$.

\section{Particular solutions for a special choice of the conformal factor
}

We consider conformally invariant modified Weyl gravity nonminimally coupled to spinor matter. This means that there is a whole class of metrics
satisfying the field equations and related by conformal transformations. It is evident that such metrics may have a different physical meaning,
although from the point of view of Weyl gravity under consideration, they all are the same solution.
Therefore, in this section we show that the class of solutions obtained by us contains the metrics which are conformally related to each other and describe physically distinct situations.

\subsection{Wormhole solution}
\label{worm_sol}

It is particularly interesting that the family of solutions obtained above contains also a wormhole solution.
To show this, consider the metric with the conformal factor
$$
  f(t, \chi ,\theta ,\varphi ) = \frac{1}{\sin \theta}.
$$
Then, introducing the new coordinate $dx = - d \theta / \sin \theta$, i.e.,
$x = \ln \cot \left( \theta / 2\right) $, the metric \eqref{solution_10} takes the form
\begin{equation}
  ds^2 = \cosh^2 x dt^2 - \frac{r^2}{4} \left\lbrace
    d x^2 + \cosh^2 x \left[
      \left( d \chi - \tanh x d \varphi \right)^2 +
      \frac{1}{\cosh^2 x} d \varphi^2
    \right]
  \right\rbrace ,
\label{4_20}
\end{equation}
and the coordinate $x$ covers the range $-\infty < x < + \infty$.  The area of a two-dimensional torus spanned on the coordinates $\chi, \varphi$
[the two-dimensional metric of this torus is given by the square brackets in Eq.~\eqref{4_20}]
goes from infinity when $x \to - \infty$, reaches its minimum for $x = 0$, and then again becomes infinite when $x \to + \infty$.

In this system of coordinates, the solution \eqref{3_a_10} takes the form
\begin{equation}
	\left( \Theta_{1,3} \right)_{m,n} =
	\pm \left( \Theta_{2,4} \right)_{m,n} =
	C \left(
		\frac{e^{-x}}{e^x + e^{- x}}
	\right)^{\alpha/2}
\left(
		\frac{e^{x}}{e^x + e^{- x}}
	\right)^{\beta/2}.
\label{4_30}
\end{equation}
For positive values of $\alpha$ and $\beta$, this expression gives soliton-like solutions.

In the case of $\alpha = \beta$, we have the symmetric solution
\begin{equation}
	\left( \Theta_{1,3} \right)_{m,n} =
	\pm \left( \Theta_{2,4} \right)_{m,n} =
	\tilde C \sech^{\alpha} x .
\label{4_40}
\end{equation}

If, say, $\beta = 0$, we get the kink-like solution (for positive $\alpha$)
\begin{equation}
	\left( \Theta_{1,3} \right)_{m,n} =
	\pm \left( \Theta_{2,4} \right)_{m,n} =
	\tilde C \left(
		\frac{e^{-x}}{e^x + e^{- x}}
	\right)^{\alpha/2}
\label{4_50}
\end{equation}
with the following asymptotic behavior:
\begin{equation}
	\left( \Theta_{1,3} \right)_{m,n} =
	\pm \left( \Theta_{2,4} \right)_{m,n}	\rightarrow
	 \begin{cases}
		0& \text{when }x \rightarrow + \infty \\
		\tilde C 				& \text{when }x \rightarrow - \infty
	\end{cases} .
\label{4_60}
\end{equation}
In the above expressions, $\tilde C$ is a rescaled integration constant.

Asymptotically, as $x\to \pm \infty$, the Ricci scalar curvature of spacetime described by the metric  \eqref{4_20} is 
$$
R=\frac{6}{r^2} \left(4-\sech^2 x\right)\to\frac{24}{r^2},
$$
i.e., there is spacetime of constant positive curvature (the asymptotically anti-de Sitter spacetime).

This solution in $F\left( B^2\right)$ modified Weyl gravity is interesting in that it describes a wormhole which is filled with a spinor field and which has a toroidal cross-section.
Such a wormhole exists in the region where quantum gravitational effects are important. When one moves away from this region, the metric \eqref{4_20}
must be matched with the corresponding GR solution.

Let us compare this solution with the known wormhole solutions whose cross-section is also toroidal.
Apparently, this type of wormholes was first considered by P.~F.~Gonzalez-Diaz in Ref.~\cite{GonzalezDiaz:1996sr}
where he studied wormholes with a cross-section,  distinct from a sphere.
In Ref.~\cite{Dzhunushaliev:2019qze}, a toroidal wormhole was considered using the thin-shell formalism by matching two
copies of flat Minkowski spacetimes along two-tori. The main feature of the solution obtained in the present paper is that it describes a wormhole filled
with quantum matter in the form of a spinor field obeying the Dirac equation.

It is appropriate to mention here that Refs.~\cite{Maldacena:2013xja,Jensen:2013ora,Susskind:2017nto}
offer an idea that entangled electrons are connected by a virtual wormhole. Of course, to confirm such a hypothesis, it
would be nice to have solutions within some kind of gravity theory. The solution obtained in the present work contains a spinor field
filling a wormhole, and hence it could be an appropriate candidate for the role of a virtual wormhole between the entangled electrons.

Notice also that to create the above wormhole the presence of exotic matter is not necessary. Apparently, this is because the wormhole was obtained not within GR but in modified gravity; this agrees with the results found in Refs.~\cite{Lobo:2008zu,Harko:2013yb}. In turn, within Einstein-Cartan gravity, it is also possible to get wormhole solutions without involving exotic matter~\cite{Bronnikov:2016xvj}.

\subsection{Cosmological solution with a bounce and inflation}

Consider now a solution where the conformal factor is chosen in the form
$$	f(t, \chi ,\theta ,\varphi ) = \frac{1}{\cos \left( t/t_0 \right)}.$$
Then, introducing the time coordinate
$d \tau = dt / \cos (t / t_0)$, that is,
$\tau = t_0 \arctanh \sin \left( t/t_0 \right)$,
we have the following metric:
$$  ds^2 = d \tau^2 - \frac{r^2}{4}
  \cosh^2 \left( \frac{\tau}{t_0} \right)
  \left[
		\left( d \chi - \cos\theta d \varphi\right)^2
		+ d \theta^2 + \sin^2 \theta d \varphi^2
  \right] .$$
It is evident that this metric describes a bounce for  $\tau = 0$ and an exponential expansion for $\tau \gg t_0$.

\subsection{Solution with a change in metric signature
}

Consider the metric
\begin{equation}
\begin{split}
  ds^2 = & d \tau^2 - \frac{r^2}{4} h(\tau)
  \left[
		\left( d \chi - \cos\theta d \varphi\right)^2
		+ d \theta^2 + \sin^2 \theta d \varphi^2
  \right]
 \\
  &
  =h(\tau) \left\lbrace
  	\frac{d \tau^2}{h(\tau)} - \frac{r^2}{4}
 	  \left[
 			\left( d \chi - \cos\theta d \varphi\right)^2
 			+ d \theta^2 + \sin^2 \theta d \varphi^2
 	  \right]
  \right\rbrace ,
\label{4_C_10}
\end{split}
\end{equation}
which, in terms of a new time coordinate $dt = d \tau / \sqrt{h}$, takes the form
$$	ds^2 = f^2 \left(
		t, \chi, \theta, \varphi
	\right) \left\{
		dt^2 - \frac{r^2}{4}
		\left[
			\left( d \chi - \cos\theta d \varphi\right)^2
			+ d \theta^2 + \sin^2 \theta d \varphi^2
		\right]
	\right\} .$$

The metric \eqref{4_C_10} is interesting in that if the function $h$ is so chosen that it changes its sign  for some $\tau_0$, for example, as
$$	h = \frac{\tau^2}{\tau_0^2} - 1 ,$$
then for $\tau < \tau_0$ the metric \eqref{4_C_10} is Euclidean, and for
 $\tau > \tau_0$ it has Lorentzian signature.
Of course, when $\tau = \tau_0$, this metric has a singularity; it is easily seen from the behavior of the scalar curvature,
$$	R \sim \frac{1}{h(\tau_0)} \to \infty .$$

An interesting fact is that the Bach and Weyl tensors remain, however, equal to zero:
$
	B_{\mu \nu}(\tau_0) = C_{\mu \nu \rho \sigma}(\tau_0) = 0
$. This means that the corresponding scalar invariants are also equal to zero:
$
	B_{\mu \nu}(\tau_0) B^{\mu \nu}(\tau_0) =
	C_{\mu \nu \rho \sigma}(\tau_0) C^{\mu \nu \rho \sigma}(\tau_0) = 0
$. The metric~\eqref{4_C_10} therefore satisfies the equations both of Weyl theory and of $F(B^2)$ modified Weyl gravity for
$\tau < \tau_0, \tau = \tau_0$, and $\tau > \tau_0$, despite the fact that at the point $\tau = \tau_0$  there is a singularity in the scalar invariants
$R, R_{\mu \nu} R^{\mu \nu}$, etc. In a certain sense, one can say that Weyl gravity `does not see' some singularities.
Apparently, this is because the Bach and Weyl tensors are conformally invariant and they do not depend on a scale of distance.

It should be also noted here that, in the Euclidean region,  it is necessary to employ the Euclidean version of the Dirac equation.
Then, for example, the Dirac matrices must satisfy the following relation:
$$	\gamma^\mu \gamma^\nu + \gamma^\nu \gamma^\mu =
	2 \delta^{\mu \nu},$$
where $\delta^{\mu \nu} $ is the Euclidean metric. The spinor  $\psi$ must in turn be transformed somehow.

From the physical point of view, this means that \textit{
in conformal Weyl gravity, as well as in $F(B^2)$ modified Weyl gravity, a change in metric signature is possible.
} The physical reason for such a change is that these theories possibly describe quantum gravitational effects which are the true reason for the change in metric signature.

The idea of a change in metric signature in cosmology was introduced in Refs.~\cite{Hartle:1983ai,Hawking:1983hj,Sakharov:1984ir}
where the authors attempt to `get rid' of a cosmological singularity at the beginning of the Universe:
for $t = 0$, spacetime is Euclidean with the metric's signature $\left( +,+,+,+\right)$;
this means that there is no singularity at this time. Then, for some value of the Euclidean time coordinate
$it = \tilde t = \tilde t_0$, there occurs a change in metric signature, and spacetime becomes Lorentzian with the metric's signature
$\left( +,-,-,-\right)$.

\section{Spin operator in the Hopf coordinates}
\label{spin_oper}

Using the triad $e^a_{\phantom{a} i}$, one can write the spin vector $\hat S_i$
in a system of the Hopf coordinates in terms of the spin vector $\hat S_a$ in a locally flat system of coordinates,
$$
	\hat S_i = e^a_{\phantom{a} i} \hat S_a  \quad \text{with} \quad
	i = \chi, \theta, \varphi ,
$$
where we use the standard spin operators
$$
	\hat S_a =	\begin{pmatrix}
		\sigma_i	&	0	\\
		0				&	\sigma_i
	\end{pmatrix},
$$
and $\sigma_i$ are the Pauli matrices. After due calculations, the spin operator takes the form
$$
	\hat S_i = f \left( t, \chi, \theta, \varphi \right)
	\begin{pmatrix}
  	\tilde \sigma_i	&	0	\\
  	0				&	\tilde \sigma_i
	\end{pmatrix},
$$
where the Pauli matrices in the Hopf coordinates are
$$
	\tilde \sigma_\chi =
	\begin{pmatrix}
		0	&	1	\\
		1	&	0	&	\\
	\end{pmatrix} ,
\quad
	\tilde \sigma_\theta =
	\begin{pmatrix}
		0	&	- i	\\
		i	&	0	\\
	\end{pmatrix} ,
\quad
	\tilde \sigma_\varphi =
	\begin{pmatrix}
		\sin \theta		&	- \cos \theta	\\
		- \cos \theta&	- \sin \theta	\\
	\end{pmatrix} .
$$

\section{Discussion and conclusions}
\label{concl}

We suggest the following physical interpretation of $F\left(B^2\right)$ modified Weyl gravity proposed here.
In a region where gravitational fields are strong, quantum GR can be approximately described by classical modified Weyl theory presented here.
Such an approximate description is analogous to what is done in Refs.~\cite{Dzhunushaliev:2013nea,Dzhunushaliev:2015mva,Liu:2016qfx}
where $F(R)$ modified gravities are regarded as an approximate description of quantum gravity in some physical processes in regions with
a strong gravitational field.

One might wonder how the transition from modified Weyl gravity (which approximately describes quantum gravitational effects) to GR occurs?
We suppose that this should happen when the magnitude of the scalar invariants becomes
$\sim \Lambda^n_{\text{QG}}$, where
$\Lambda_{\text{QG}} = l^{-1}_{QG}$ is some constant which has the dimension of an inverse length and
$n$ is an integer which provides the necessary dimension of any given scalar invariant. It seems to us that, in some sense, this process
is analogous to the appearance of the quantity  $\Lambda_{\text{QCD}}$ with the dimension of an inverse length in conformally invariant Yang-Mills theory (in QCD).
In QCD, such a quantity separates a region where the use of perturbative calculations is possible from a region where one has to employ nonperturbative computations.
For example, one can choose $F \left( B^2 \right) = \Lambda_{\text{QG}}^{-8} B^2$.

The next question to be addressed concerns the magnitude of the dimensional constant $\Lambda_{\text{QG}}$. It seems to us that
$\Lambda_{\text{QG}} \neq l^{-1}_{\text{Pl}}$. The reason is that the magnitude of this quantity must occur from the nonperturbative quantization of GR.
If we compare our situation with QCD, the latter involves the quantity $\Lambda_{\text{QCD}}$ that cannot be obtained from constants appearing in the QCD Lagrangian.
In our case we also suppose that $\Lambda_{\text{QG}}$ is not expressed in terms of constants appearing in the GR Lagrangian. This means that $\Lambda_{\text{QG}}$
cannot be derived from such constants like the velocity of light $c$, the Planck constant $\hbar$, and the Newtonian gravitational constant  $G_N$,
i.e., from the Planck length $l_{\text{Pl}} = \left(\hbar G_N / c^3\right)^{1/2}$.

Note that if there exists the dimensional parameter $\Lambda_{\text{QG}}$ which separates a region where
the scalar curvature $R \gg \Lambda_{\text{QG}}^2$ and one has to take into account quantum gravitational effects from a region where
$R \ll \Lambda_{\text{QG}}^2$ and the gravitational field is described by classical GR, the dimensionless parameter
 $l_{Pl}/l_{\text{QG}}$ occurs, which allows to take into account quantum gravitational effects perturbatively. For example, using this quantity, one can in
principle calculate the corrections to the solution obtained here within Weyl gravity. These corrections can already go beyond the framework of conformal Weyl gravity and
take into account a contribution coming from off-diagonal metric components describing rotation/spin of matter.

Notice also the following unexpected feature of the solutions obtained. The Dirac equation under consideration is linear, and it describes free particles, i.e.,
there are no any confining external potentials.  For this reason, the appearance of the discrete spectrum of regular solutions (for which $\int \left| \psi \right|^2 d V < \infty$)
is absolutely unexpected fact. This fact, at least in two cases, can be explained as follows. In the case of
$f \left( \chi, \theta, \varphi \right) = 1$, the three-dimensional cross-section is a Hopf sphere, and the presence of the nontrivial solution for the spinor field is caused by
the fact that, for this solution, there exists a nonzero value of the Hopf invariant~\eqref{A_150}.
In the case when the conformal factor
$
  f \left( t, \chi, \theta, \varphi \right) =
  r^2\left[
  \sin^2 \left( \frac{\theta + \varphi - \chi - \pi}{4} \right) +
  \sin^2 \left( \frac{\theta - \varphi + \chi + \pi}{4} \right)
  \right]^{-2}
$ [see Eq.~\eqref{A_140}],
the three-dimensional cross-section is a flat Euclidean space, but with the nontrivial flow of time   according to the expression
$
 d \tau = f \left( \chi, \theta, \varphi \right) dt
$, where $\tau$ is the proper time of an observer at a given point of spacetime; this results in the appearance of nontrivial solutions to the Dirac equation.

In modified Weyl gravity suggested in the present paper, quantum gravitational effects in strong gravitational fields have the result that physical processes
(perhaps not always but only in some cases) become independent on the scale.

One more conclusion, which follows from the present study, is that (as it seems to us) an exact quantization of GR must be performed such that in some situations a quantized gravitational field
could be approximately described by some modified gravities. For example, a physical system with a strong quantized gravitational field interacting with a spinor
field can be approximately described by modified Weyl gravity.

Thus, in the present paper:
\begin{itemize}
\item We have suggested modified Weyl gravity where a nonminimal interaction between matter and gravity is introduced.
\item As a source of matter, we have used a massless spinor field described by one Dirac spinor. For such a self-consistent system, we have obtained a solution for the gravitating spinor field for the metrics which are conformally equivalent to the Hopf metric on the Hopf fibration.
\item It is pointed out that, using the standard Gram-Schmidt process, one can orthogonalize the eigenwave functions of the Dirac equation.
\item We show that it is possible to relate the current density of the solution obtained for the spinor field to the Hopf invariant.
\item The expressions for the spin operators in the Hopf coordinates have been obtained.
\item It is shown that, within the class of conformally equivalent metrics under investigation, there are:
(a) a metric describing a toroidal wormhole supported by ordinary spinor field without involving exotic matter;
(b) a metric describing a bounce of the Universe and its subsequent inflation;
and (c) a transition between metrics with different signatures.
\item The physical interpretation of $F\left( B^2 \right) $ modified Weyl gravity involving a massless Dirac spinor field is suggested.
\end{itemize}

In conclusion, we emphasize  that $F\left(B^2\right)$  modified Weyl gravity studied in the present paper should not be regarded as a fundamental theory,
since, in our opinion, in some physical situations, it can be viewed only as some effective description of quantum GR in regions where gravitational fields are strong.

\section*{Acknowledgments}
We gratefully acknowledge support provided by Grant No.~BR05236494
in Fundamental Research in Natural Sciences by the Ministry of Education and Science of the Republic of Kazakhstan. We are also grateful
to the Research Group Linkage Programme of the Alexander von Humboldt Foundation for the support of this research.

\appendix

\section{Hopf fibration. Stereographic projection}
\label{Hopf_bundle}

The standard three-dimensional sphere $S^3$ is a set of points in $\mathbb{R}^4$  that satisfy the equation
$$
X^2 + Y^2 + Z^2 + W^2 = r^2,
$$
where $r$ is the radius of the sphere and $X, Y, Z, W$ are coordinates of the enveloping four-dimensional space in which the three-dimensional sphere is embedded.

The fibration $h: S^3 \rightarrow S^2$ is defined by the Hopf mapping
$$
h \left( X, Y, Z, W \right) = \left\{
X^2 + Y^2 - Z^2 - W^2,
2 \left( X W + Y Z \right) ,
2 \left( Y W - X Z \right)
\right\}.
$$
The Hopf coordinates  $0 \leq \theta \leq \pi$, $0 \leq \chi, \varphi \leq 2 \pi$ on a three-dimensional sphere are given in the form
\begin{eqnarray}
	X &=& r \cos \left( \frac{\varphi + \chi}{2} \right) \sin \frac{\theta}{2},
\quad
	Y = r \sin \left( \frac{\varphi + \chi}{2} \right) \sin \frac{\theta}{2},
\label{A_40}\nonumber\\
	Z &=& r \cos \left( \frac{\varphi - \chi}{2} \right) \cos \frac{\theta}{2},
\quad
	W = r \sin \left( \frac{\varphi - \chi}{2} \right) \cos \frac{\theta}{2}.
\label{A_60}\nonumber
\end{eqnarray}
The stereographic projection of a three-dimensional sphere embedded into a four-dimensional space with the coordinates $X, Y, Z, W$
on a three-dimensional space with the coordinates $x, y, z$ is defined by the formulas
\begin{equation}
	x = r \frac{X}{r - W},
\quad
	y = r \frac{Y}{r - W},
\quad
	z = r \frac{Z}{r - W},
\label{A_90}
\end{equation}
where $r$ is the radius of the three-dimensional sphere.

For the stereographic projection \eqref{A_90}, the coordinates $x, y, z$ are expressed in terms of the Hopf coordinates on a sphere as
\begin{equation}
	x = r \frac{
		\sin \left( \frac{\theta}{2} \right)
		\cos \left(\frac{\varphi + \chi}{2}\right)
	}
	{1 - \cos \left( \frac{\theta}{2} \right) \sin \left(\frac{\varphi - \chi}{2}\right) },
\quad
	y = r \frac{
		\sin \left( \frac{\theta}{2} \right)
		\sin \left(\frac{\varphi + \chi}{2}\right)
	}
	{1 - \cos \left( \frac{\theta}{2} \right) \sin \left(\frac{\varphi - \chi}{2}\right) },
	\quad
	z = r \frac{
		\cos \left( \frac{\theta}{2} \right)
		\cos \left(\frac{\varphi - \chi}{2}\right)
	}
	{1 - \cos \left( \frac{\theta}{2} \right) \sin \left(\frac{\varphi - \chi}{2}\right) } .
\label{A_120}
\end{equation}
The Hopf metric on a three-dimensional sphere is
\begin{equation}
	dl^2 = dX^2 + dY^2 + dZ^2 + dW^2 = \frac{r^2}{4}
	\left[
		\left( d \chi - \cos\theta d \varphi\right)^2
		+ d \theta^2 + \sin^2 \theta d \varphi^2
	\right] = r^2 dS^2_3 ,
\label{A_130}
\end{equation}
where $dS^2_3$ is the metric on a unit three-dimensional sphere. Using the stereographic projection~\eqref{A_120},
the metric in the three-dimensional space  $x, y, z$ with the Hopf coordinates $\chi, \theta, \varphi$ takes the form
\begin{equation}
\begin{split}
	dl^2 = & \frac{r^2}{\left[
		2 - \sin \left( \frac{\theta + \varphi - \chi}{2} \right) +
		\sin \left( \frac{\theta - \varphi + \chi}{2} \right)
		\right]^2 }
	\left[
	\left( d \chi - \cos\theta d \varphi\right)^2
	+ d \theta^2 + \sin^2 \theta d \varphi^2
	\right]
	\\
	&=\omega_\chi^2 + \omega_\theta^2 + \omega_\varphi^2 =
	\frac{4 r^2}{\left[
		2 - \sin \left( \frac{\theta + \varphi - \chi}{2} \right) +
		\sin \left( \frac{\theta - \varphi + \chi}{2} \right)
		\right]^2 } dS^2_3 =
	\frac{r^2}{\left[
		\sin^2 \left( \frac{\theta + \varphi - \chi - \pi}{4} \right) +
		\sin^2 \left( \frac{\theta - \varphi + \chi + \pi}{4} \right)
		\right]^2 } dS^2_3 .
\end{split}
\label{A_140}
\end{equation}
That is, this metric is conformally equivalent to the Hopf metric on a three-dimensional sphere. We will use the Hopf coordinates and the metric \eqref{A_140} in obtaining the solutions to the Dirac equation.

The transition matrix $R$ between the 1-forms $dx, dy,dz$ and
$\omega_\chi, \omega_\theta, \omega_\varphi$
\begin{equation}
  \begin{pmatrix}
    dx  \\  dy  \\  dz
  \end{pmatrix} = R
   \begin{pmatrix}
    \omega_\chi  \\  \omega_\theta  \\  \omega_\varphi
  \end{pmatrix},
\label{A_141}
\end{equation}
which is a spatial part of the Lorentz transformation matrix
$$
  \Lambda^a_{\phantom{a} b} =
  \begin{pmatrix}
    1 & 0 \\
    0 & R
  \end{pmatrix} ,
$$
has the form
{\footnotesize
\begin{equation}
\begin{split}
  & R =
  \frac{r}{4
    \left[
    \cos \left(\frac{\theta }{2}\right)
    \sin \left(\frac{\varphi -\chi }{2}\right)-1
    \right]
  }
\\
&
  \begin{pmatrix}
    2 \sin \left(\frac{\theta }{2}\right) \left[
      \cos \left(\frac{\theta }{2}\right) \cos \varphi +
      \sin \left(\frac{\varphi +\chi }{2}\right)
    \right] &
    - 2 \cos \left(\frac{\varphi +\chi }{2}\right) \left[
      \cos \left(\frac{\theta }{2}\right) -
      \sin \left(\frac{\varphi -\chi }{2}\right)
    \right] &
    4 \cos \left(\frac{\theta }{2}\right)
    \sin \left(\frac{\varphi + \chi }{2}\right) +
    2 \cos (\theta) \cos (\varphi ) -
    2 \cos (\chi )
  \\
    - 2 \sin \left(\frac{\theta }{2}\right) \left[
      \cos \left(\frac{\varphi +\chi }{2}\right) -
      \cos \left(\frac{\theta }{2}\right) \sin (\varphi )
    \right] &
    - 2 \sin \left(\frac{\varphi +\chi }{2}\right) \left[
      \cos \left(\frac{\theta }{2}\right) -
      \sin \left(\frac{\varphi -\chi }{2}\right)
    \right] &
    4 \cos \left(\frac{\theta}{2}\right)
    \cos \left(\frac{\varphi + \chi }{2} \right) -
    2 \cos (\theta ) \sin (\varphi ) + 2 \sin (\chi )
  \\
    2 \cos \left(\frac{\theta }{2}\right) \left[
    \cos \left(\frac{\theta }{2}\right) -
    \sin \left(\frac{\varphi -\chi }{2}\right)
    \right] &
    2 \sin \left(\frac{\theta }{2}\right)
    \cos \left(\frac{\varphi -\chi }{2}\right)  &
    - 2 \sin \left(\frac{\theta }{2}\right) \left[
        \cos \left(\frac{\theta }{2}\right) -
        \sin \left(\frac{\varphi -\chi }{2}\right)
    \right]
  \end{pmatrix} .
\end{split}
\label{A_143}\nonumber
\end{equation}
}

After the transformation \eqref{A_141}, the spinor $\psi_H$ written in the Hopf coordinates takes the form
$$
  \psi_D = S \psi_H,
$$
where $S$ is the matrix of an SU(2) transformation defined from the expression
$$
  S^{-1} \gamma^a S = \Lambda^a_{\phantom{a} b} \gamma^b.
$$

On the Hopf fibration, one can introduce the so-called Hopf invariant, which is defined as
\begin{equation}
	H = \frac{1}{V} \int \Upsilon \wedge d \Upsilon ,
\label{A_150}
\end{equation}
where $\Upsilon$ is some 1-form and the volume of a three-dimensional Hopf sphere is
$V = \int \sin \theta d \chi d \theta d \varphi$. On the Hopf fibration, the 1-form
 \begin{equation}
	\Upsilon = d \chi - \cos\theta d \varphi .
\label{A_160}
\end{equation}
For this case, the Hopf invariant
$
  H = 1 .
$


\begin{thebibliography}{99}

\bibitem{Finster:1998ws}
F.~Finster, J.~Smoller, and S.~T.~Yau,
Phys.\ Rev.\ D {\bf 59}, 104020 (1999).

\bibitem{Finster:1998ux}
F.~Finster, J.~Smoller, and S.~T.~Yau,
Phys.\ Lett.\ A {\bf 259}, 431 (1999).

\bibitem{Herdeiro:2017fhv}
C.~A.~R.~Herdeiro, A.~M.~Pombo, and E.~Radu,
Phys.\ Lett.\ B {\bf 773}, 654 (2017).

\bibitem{Dzhunushaliev:2018jhj}
  V.~Dzhunushaliev and V.~Folomeev,
  Phys.\ Rev.\ D {\bf 99}, no. 8, 084030 (2019).

\bibitem{Dzhunushaliev:2019kiy}
  V.~Dzhunushaliev and V.~Folomeev,
  Phys.\ Rev.\ D {\bf 99}, no. 10, 104066 (2019).

\bibitem{Sushkov:2009hk}
  S.~V.~Sushkov,
  Phys.\ Rev.\ D {\bf 80}, 103505 (2009).

\bibitem{Tamanini:2013aca}
  N.~Tamanini and T.~S.~Koivisto,
  Phys.\ Rev.\ D {\bf 88}, no. 6, 064052 (2013).

\bibitem{Dzhunushaliev:2013nea}
V.~Dzhunushaliev, V.~Folomeev, B.~Kleihaus, and J.~Kunz,
  Eur.\ Phys.\ J.\ C {\bf 74}, 2743 (2014).

\bibitem{Harko:2014gwa}
  T.~Harko and F.~S.~N.~Lobo,
  Galaxies {\bf 2}, no. 3, 410 (2014).

\bibitem{Dzhunushaliev:2015mva}
 V.~Dzhunushaliev, V.~Folomeev, B.~Kleihaus, and J.~Kunz,
  Eur.\ Phys.\ J.\ C {\bf 75}, no. 4, 157 (2015).

\bibitem{Nojiri:2010wj}
  S.~Nojiri and S.~D.~Odintsov,
  Phys.\ Rept.\  {\bf 505}, 59 (2011).

\bibitem{Nojiri:2017ncd}
  S.~Nojiri, S.~D.~Odintsov, and V.~K.~Oikonomou,
  Phys.\ Rep.\  {\bf 692}, 1 (2017).

\bibitem{Mannheim:2009qi}
  P.~D.~Mannheim,
  Gen.\ Rel.\ Grav.\  {\bf 43}, 703 (2011).

\bibitem{Flanagan:2006ra}
  E.~E.~Flanagan,
  Phys.\ Rev.\ D {\bf 74}, 023002 (2006).

\bibitem{Mannheim:1988dj}
  P.~D.~Mannheim and D.~Kazanas,
  Astrophys.\ J.\  {\bf 342}, 635 (1989).

\bibitem{Mannheim:2010xw}
  P.~D.~Mannheim and J.~G.~O'Brien,
  Phys.\ Rev.\ D {\bf 85}, 124020 (2012).

\bibitem{Mannheim:2005bfa}
  P.~D.~Mannheim,
  Prog.\ Part.\ Nucl.\ Phys.\  {\bf 56}, 340 (2006).

\bibitem{Diaferio:2011kc}
  A.~Diaferio, L.~Ostorero, and V.~F.~Cardone,
  JCAP {\bf 1110}, 008 (2011).

\bibitem{Lawrie2002}
I.~Lawrie, {\it A unified grand tour of theoretical physics} (Institute of Physics
Publishing, Bristol and Philadelphia, 2002).

\bibitem{Bronnikov:2019nqa}
K.~A.~Bronnikov, Y.~P.~Rybakov, and B.~Saha,
  Eur.\ Phys.\ J.\ Plus {\bf 135}, no. 1, 124 (2020).


\bibitem{GonzalezDiaz:1996sr}
P.~F.~Gonzalez-Diaz,
Phys.\ Rev.\ D {\bf 54}, 6122 (1996).

\bibitem{Dzhunushaliev:2019qze}
 V.~Dzhunushaliev, V.~Folomeev, B.~Kleihaus, and J.~Kunz,
  Phys.\ Rev.\ D {\bf 99}, no. 4, 044031 (2019).

\bibitem{Maldacena:2013xja}
 J.~Maldacena and L.~Susskind,
  Fortsch.\ Phys.\  {\bf 61}, 781 (2013).

\bibitem{Jensen:2013ora}
 K.~Jensen and A.~Karch,
  Phys.\ Rev.\ Lett.\  {\bf 111}, no. 21, 211602 (2013).

\bibitem{Susskind:2017nto}
 L.~Susskind and Y.~Zhao,
  Phys.\ Rev.\ D {\bf 98}, no. 4, 046016 (2018).

\bibitem{Lobo:2008zu}
  F.~S.~N.~Lobo,
  Class.\ Quant.\ Grav.\  {\bf 25}, 175006 (2008).

\bibitem{Harko:2013yb}
T.~Harko, F.~S.~N.~Lobo, M.~K.~Mak, and S.~V.~Sushkov,
Phys.\ Rev.\ D {\bf 87}, no. 6, 067504 (2013).

\bibitem{Bronnikov:2016xvj}
K.~A.~Bronnikov and A.~M.~Galiakhmetov,
  Phys.\ Rev.\ D {\bf 94}, no. 12, 124006 (2016).

\bibitem{Hartle:1983ai}
  J.~B.~Hartle and S.~W.~Hawking,
  Phys.\ Rev.\ D {\bf 28}, 2960 (1983)
  [Adv.\ Ser.\ Astrophys.\ Cosmol.\  {\bf 3}, 174 (1987)].

\bibitem{Hawking:1983hj}
  S.~W.~Hawking,
  Nucl.\ Phys.\ B {\bf 239} (1984) 257
   [Adv.\ Ser.\ Astrophys.\ Cosmol.\  {\bf 3} (1987) 236].

\bibitem{Sakharov:1984ir}
  A.~D.~Sakharov,
  Sov.\ Phys.\ JETP {\bf 60}, 214 (1984)
  [Zh.\ Eksp.\ Teor.\ Fiz.\  {\bf 87}, 375 (1984)]
  [Sov.\ Phys.\ Usp.\  {\bf 34}, 409 (1991)]
  [Usp.\ Fiz.\ Nauk {\bf 161}, no. 5, 94 (1991)].

\bibitem{Liu:2016qfx}
 X.~Liu, T.~Harko, and S.~D.~Liang,
  Eur.\ Phys.\ J.\ C {\bf 76}, no. 8, 420 (2016).

\end{thebibliography}
\end{document}